\documentstyle[12pt]{article}

\makeatletter
\renewcommand\thesubsection{\thesection.\@arabic\c@subsection}
\makeatother

\newcommand{\sect}[1]{\setcounter{equation}{0}\section{#1}}

\newcommand {\beq}{\begin{equation}}
\newcommand {\eeq}{\end{equation}}
\newcommand {\beqa}{\begin{eqnarray}}
\newcommand {\eeqa}{\end{eqnarray}}         
\newcommand {\beqs}{\begin{eqnarray*}}
\newcommand {\eeqs}{\end{eqnarray*}}
\newcommand {\bds}{\begin{displaymath}}
\newcommand {\eds}{\end{displaymath}}
\newcommand {\n}{\nonumber\\}

\newcommand{\no}{\noindent}

\newcommand {\bebb}{\begin{thebibliography}}
\newcommand {\eebb}{\end{thebibliography}}      
\newcommand {\bbit}{\bibitem}

\def\a{\alpha}
\def\bt{\beta}

\def\G{\Gamma}
\def\gm{\gamma}

\def\tl{\tilde}  

\def\p{\partial}

\def\lt{\left}
\def\rt{\right}

\def\rtarr{\rightarrow}

\def\dg{\dagger}

\def\ph{\phi}
\def\ps{\psi}
\def\psd{\psi ^{\dagger}}

\def\psb{\bar{\psi}}
\def\psbd{\bar{\psi} ^{\dagger}}

\def\journal#1&#2(#3){\unskip, \sl #1\ \bf #2 \rm(19#3) }
\def\andjournal#1&#2(#3){\sl #1~\bf #2 \rm (19#3) }

\def\npb#1#2#3{Nucl. Phys. {\bf B#1}, (#2) #3}

\def\plb#1#2#3{Phys. Lett. {\bf B#1}, (#2) #3}

\def\jmp#1#2#3{J. Math. Phys. {\bf #1}, (#2) #3}

\begin{document}


\begin{flushright}
\end{flushright}

\vskip 1cm

\begin{center}
{\large\bf Coherent State Construction of Representations of $osp(2|2)$
and Primary Fields of $osp(2|2)$ Conformal Field Theory }

\vspace{1cm}

{\large  Yao-Zhong Zhang }
\vskip.1in
{\em  Department of Mathematics, 
University of Queensland, Brisbane, Qld 4072, Australia}

\end{center}

\date{}



\begin{abstract}

Representations of the superalgebra $osp(2|2)$ and current superalgebra
$osp(2|2)^{(1)}_k$ in the standard basis are investigated. All 
finite-dimensional typical and atypical representations of $osp(2|2)$ 
are constructed by the vector coherent state method. Primary fields of
the non-unitary conformal field theory associated with $osp(2|2)^{(1)}_k$
in the standard basis are obtained for arbitrary level $k$.

\end{abstract}



\vspace{0.5cm}


\setcounter{section}{0}
\setcounter{equation}{0}
\sect{Introduction}

Superalgebras and their corresponding non-unitary conformal field
theories (CFTs) have recently attracted much interests in high energy and
condensed matter physics communities, partly because of their
applications in areas such as topological field theory \cite{Roz92,Isi94},
logarithmic CFTs (see e.g. \cite{Flohr01} and references therein)
and disordered systems \cite{Efe83,Ber95,Mud96,Maa97,Bas00,Gur00}. There the 
vanishing of superdimensions and Virasoro central charges and the
existence of primary fields with negative dimensions are essential
\cite{Ber95,Mud96,Lud00,Bha01}. The most
interesting superalgebras satisfying these requirements are $osp(N|N)$ 
and $gl(N|N)$. 

It is well-known that unlike a purely bosonic algebra 
a superalgebra admits different Weyl inequivalent
choices of simple root systems, which correspond to inequivalent Dynkin 
diagrams. In the case of $osp(2|2)$, one has two choices of simple root
systems which are unrelated by Weyl transformations: a system of fermionic and
bosonic simple roots (i.e. the so-called standard basis), or a purely
fermionic system of simple roots (that is the so-called non-standard
basis). So it is desirable to obtain results in the two different bases
for different physical applications. 

Representations of $osp(2|2)$ have been constructed only in the non-standard
basis \cite{Sch77,Mar80} (see also \cite{Dob86,Kob94,Bow97,Sem97} for the
study of certain specific representations). 
Primary fields of the $osp(2|2)$ non-unitary CFT
have been investigated in \cite{Ding03}. However, for the standard basis
case the expressions found in \cite{Ding03} are valid only for a class of
atypical representations corresponding to $q=p$ (see section 4 below).

In this paper we construct explicitly all finite-dimensional 
typical and atypical representations of $osp(2|2)$ in the
standard basis by using the vector coherent state method. We moreover
construct all primary fields in the standard basis
for the $osp(2|2)$ non-unitary CFT. The operator product expansions (OPEs)
of the $osp(2|2)$ currents with the primary fields are presented.
Our results are expected to be useful in the supersymmetric method to
certain Gaussian disordered systems.

\setcounter{section}{1}
\setcounter{equation}{0}
\sect{Boson-fermion realizations of $osp(2|2)$}

Superalgebra $osp(2|2)$ is a ${\bf Z}_2$-graded algebra, $osp(2|2)=
osp(2|2)^{\rm even}\oplus osp(2|2)^{\rm odd}$, with
\beq
osp(2|2)^{\rm even}=u(1)\oplus sl(2)=\{H'\}\oplus\{H, E, F\},~~~~
osp(2|2)^{\rm odd}=\{e,f,\bar{e},\bar{f}\},
\eeq
where $e,\;f,\;\bar{e},\;\bar{f}$ are the generators corresponding 
fermionic roots, and $E,\;F$ are those to bosonic roots.

In the standard basis, $E\;(F)$ and $e\;(f)$ are the generators 
corresponding to the even and odd simple roots of $osp(2|2)$,
respectively, and $\bar{e},\; \bar{f}$ are the odd non-simple generators. 
They satisfy the following (anti-)commutation relations: 
\beqa
 &&[E,F]=H, ~~~~[H,E]=2E,~~~~[H,F]=-2F, \n
 &&\{e,f\}=-\frac{1}{2}(H-H^{\prime}), ~~~~[H,e]=-e,~~~~[H,f]=f, \n
 &&[H^{\prime},e]=-e,~~~~[H^{\prime},f]=f,\n
 &&[E,e]=\bar{e},~~~~[F,f]=\bar{f}, \n
 &&\{\bar{e},\bar{f}\}=-\frac{1}{2}(H+H^{\prime}),~~~~
   [H,\bar{e}]=\bar{e},~~~~[H,\bar{f}]=-\bar{f},\n
 &&[H^{\prime},\bar{e}]=-\bar{e},~~~~[H^{\prime},\bar{f}]=\bar{f},\n
 &&\{e,\bar{f}\}=-F,~~~~\{\bar{e},f\}=E,~~~~ 
   [E,\bar{f}]=f,~~~~[F,\bar{e}]=e.
\eeqa

\no All other (anti-)commutators are zero. The quadratic Casimir 
is given by
\beq
C_2=\frac{1}{2}\left(H(H+2)- H^{\prime}(H^{\prime}+2)\right) 
+ 2fe -2 {\bar f}{\bar e}+2FE.
\eeq

Let $|hw>$ be the highest weight state of highest weight $(p,q)$
of $osp(2|2)$ in the standard basis:
\beq
H|hw>=2p|hw>,~~~~H'|hw>=2q|hw>,~~~~E|hw>=e|hw>=\bar{e}|hw>=0.
\eeq
Define the vector coherent states, $e^{Fa+ \a_1 f+\bar{f}\a_2}|hw>$.
As is known from \cite{Sch77}, the definition of an (grade) adjoint
action on the Lie superalgebra generators is not unique. Therefore
the order between the fermionic operators in the exponent defining the
coherent states is inmaterial in the sense that it will not affect the
conclusions of this paper. Here we have chosen a order such
that the dual state has the form appeared in (\ref{function}) below.

Then state vectors $|\psi>$ are mapped into functions
\beqa
\psi_{p,q}&=&<hw|\exp\lt(a^\dg E+\a^\dg_1 e+\a^\dg_2 \bar{e}\rt)|\psi>|0>\n
&=&<hw|e^{a^\dg E}e^{\a_1^\dg e}
   e^{(\a_2^\dg-\frac{1}{2}\a_1^\dg a^\dg)\bar{e}}|\psi>|0>.\label{function}
\eeqa
Here $a, a^\dg$ are bosonic operators with number operator $N_a$, and
$\a_1~(\a_1^\dg), \a_2~(\a_2^\dg)$ are fermionic operators with number
operators $N_{\a_1}, N_{\a_2}$, respectively. These operators satisfy
relations:
\beqa
&&[a, a^\dg]=1,~~~~[N_a, a^\dg]=a^\dg,~~~~[N_a, a]=-a,\n
&&\{\a_1,\a_1^\dg\}=1,~~~~[N_{\a_1}, \a_1^\dg]=\a_1^\dg,~~~~[N_{\a_1},
   \a_1]=-\a_1,\n
&&\{\a_2,\a_2^\dg\}=1,~~~~[N_{\a_2}, \a_2^\dg]=\a_2^\dg,~~~~[N_{\a_2},
   \a_2]=-\a_2,
\eeqa
all other (anti-)commutators are zero. Moreover,
$a|0>=\a_1|0>=\a_2|0>=0$.

Operators $A$ are mapped as follows
\beq
A|\psi>\rtarr \G(A)\psi_{J,q}=<hw|e^{a^\dg E}e^{\a_1^\dg e}
e^{(\a_2^\dg-\frac{1}{2}\a_1^\dg a^\dg)\bar{e}}A|\psi>|0>.
\eeq
Taking $A=H,H',E, \cdots$ in turn and after some algebraic manipulations,
we find
\beqa
\G(H)&=&2p-2N_a+N_{\a_1}-N_{\a_2},~~~~\G(H')=2q+N_{\a_1}+N_{\a_2},\n
\G(E)&=&a-\frac{1}{2}\a_1^\dg\a_2,\n
\G(F)&=&2pa^\dg-\a_2^\dg\a_1-a^\dg \lt(N_a-\frac{1}{2}N_{\a_1}+\frac{1}{2}
    N_{\a_2}\rt)-\frac{1}{4}(a^\dg)^2\a_1^\dg\a_2,\n
\G(e)&=&\a_1+\frac{1}{2}a^\dg \a_2,~~~~\G(f)=-(p-q)\a_1^\dg+\a_2^\dg a
   +\frac{1}{2}\a_1^\dg\lt(N_a+N_{\a_2}\rt),\n
\G(\bar{e})&=&\a_2,~~~~\G(\bar{f})=-(p+q)\a_2^\dg
   -\frac{1}{2}(3p-q)a^\dg \a_1^\dg+\a_2^\dg\lt(N_a-N_{\a_1}\rt)
   +\frac{1}{2}a^\dg\a_1^\dg N_a.\label{b-f}\n
\eeqa
This gives a free boson-fermion realization of $osp(2|2)$ in the
standard basis.  In this realization, the Casimir takes a constant
value, i.e. $C_2=2[p(p+1)-q(q+1)]$.

\sect{Construction of representations of $osp(2|2)$}

Unlike ordinary bosonic algebras, there are two types of representations
for most superalgebras. They are the so-called typical and atypical
representations. The typical representations are irreducible and are
similar to the usual representations appeared in ordinary bosonic
algebras. The atypical representations have no counterpart in the
representation theory of bosonic algebras. They can be irreducible or
not fully reducible (i.e. reducible or indecomposable). This makes the
study of representations of superalgebras difficult in general.

In this section we use the above free boson-fermion realization to construct 
finite-dimensional representations of $osp(2|2)$ in the standard bases. 
As we will see, all finite-dimensional typical and atypical
representations of $osp(2|2)$ can be constructed in an unified manner.

To begin, we note that representations of $osp(2|2)$ in the standard basis
are labelled by
$(p,q)$ with $p$ being a positive integer or half-integer and $q$ an
arbitrary complex number. There are  four independent combinations of
creation operators acting on the vacuum vector $|0>$:
\beqa
&&\lt(a^\dg\rt)^{p-m}|0>,~~~p-m\in {\bf Z}_+,\n
&&\a_1^\dg\lt(a^\dg\rt)^{p-m-1/2}|0>,~~~p-\frac{1}{2}-m\in {\bf Z}_+,\n
&&\a_2^\dg\lt(a^\dg\rt)^{p-m-3/2}|0>,~~~p-\frac{3}{2}-m\in {\bf Z}_+,\n
&&\a_1^\dg\a_2^\dg\lt(a^\dg\rt)^{p-m-2}|0>,~~~p-2-m\in {\bf Z}_+.
    \label{com-st}
\eeqa
Thus each $osp(2|2)$ representation decomposes into at most four
representations of the even subalgebra $sl(2)\oplus u(1)$. Let us
construct representations for $sl(2)\oplus u(1)$ out of the above
states. It is easy to check that the first and
the last states are already representations of $sl(2)\oplus u(1)$ with highest
weight weights $(p,q)$ and $(p,q+1)$, respectively. We denote these two
multiplets by $|p,m;q>$ and $|p,m;q+1>$, respectively. We now show
that the second and the third states can be combined into two independent
multiplets of $sl(2)\oplus u(1)$ with highest weights $(p-\frac{1}{2},
q+\frac{1}{2})$ and $(p+\frac{1}{2},q+\frac{1}{2})$, respectively.
Let
\beq
\chi_{p,q}^m=\frac{1}{2}c_{p,q}^m\a_1^\dg\lt(a^\dg\rt)^{p-m-1/2}|0>
   +\bar{c}_{p,q}^m\a_2^\dg\lt(a^\dg\rt)^{p-m-3/2}|0>,
\eeq
where $c_{p,q}^m$ and $\bar{c}_{p,q}^m$ are functions of $p,q,m$ to be
determined. Then,
\beqa
\G(E)\chi_{p,q}^m&=&\frac{1}{2}\lt((p-m-\frac{1}{2})c_{p,q}^m-\bar{c}_{p,q}^m
  \rt)\a_1^\dg\lt(a^\dg\rt)^{p-m-3/2}|0>\n
& &+(p-m-\frac{3}{2})\bar{c}_{p,q}^m\a_2^\dg\lt(a^\dg\rt)^{p-m-5/2}|0>.
\eeqa
To make the representation finite-dimensional, the r.h.s. of this
equation must equal to
$(p-m-\frac{x}{2})\chi_{p,q}^{m+1}$ for some integer $x$. This requires
\beqa
(p-m-\frac{x}{2})\bar{c}_{p,q}^{m+1}&=&(p-m-\frac{3}{2})\bar{c}_{p,q}^m,\n
(p-m-\frac{x}{2})c_{p,q}^{m+1}&=&(p-m-\frac{1}{2})c_{p,q}^m-\bar{c}_{p,q}^m.
  \label{relation1} 
\eeqa
In view of the 2nd and 3rd equations of (\ref{com-st}), the maximum
value that $m$ can have is $p-1/2$ or $p-3/2$. This means that $x$ 
can only be 1 or 3. So we have two cases to consider:\\
\underline{Case 1}: $x=1$, for this case (\ref{relation1}) has solutions
\beq
\bar{c}_{p,q}^m=(p-m-\frac{1}{2})X_{p,q},~~~~
c_{p,q}^m-c_{p,q}^{m+1}=X_{p,q}.\label{soln1-case1}
\eeq
\underline{Case 2}: $x=3$, for this case, one has from
(\ref{relation1})
\beq
\bar{c}_{p,q}^m=Y_{p,q},~~~~(p-m-\frac{1}{2})c_{p,q}^m-(p-m-\frac{3}{2})
c_{p,q}^{m+1}=Y_{p,q}.\label{soln1-case2}
\eeq
Here $X_{p,q}$ and $Y_{p,q}$ only depend on $p$ and $q$. On the other
hand,
\beqa
\G(F)\chi_{p,q}^m&=&\frac{1}{2}\lt((p+m+1)c_{p,q}^m-\frac{1}{2}
  \bar{c}_{p,q}^m\rt)\a_1^\dg\lt(a^\dg\rt)^{p-m+1/2}|0>\n
& &\lt((p+m+1)\bar{c}_{p,q}^m-\frac{1}{2}c_{p,q}^m\rt)
  \a_2^\dg\lt(a^\dg\rt)^{p-m-1/2}|0>.
\eeqa
This must equal to $(p+m+\frac{y}{2})\chi_{p,q}^{m-1}$ for some integer
$y$ in order for the representation to be finite-dimensional. So we get
\beqa
\frac{1}{2}c_{p,q}^m&=&(p+m+1)\bar{c}_{p,q}^m-(p+m+\frac{y}{2})
   \bar{c}_{p,q}^{m-1},\n
\frac{1}{2}\bar{c}_{p,q}^m&=&(p+m+1)c_{p,q}^m-(p+m+\frac{y}{2})
   c_{p,q}^{m-1}.\label{relation2}
\eeqa
Combining (\ref{relation2}) with (\ref{soln1-case1}) or
(\ref{soln1-case2}), we obtain 
\beq
y=3,~~~c_{p,q}^m=3p+m+\frac{5}{2},~~~\bar{c}_{p,q}^m=-(p-m-\frac{1}{2})
\eeq
for Case 1, and
\beq
y=1,~~~~c_{p,q}^m=\bar{c}_{p,q}^m=1
\eeq
for Case 2. Here we have set the overall factors $X_{p,q}$ and
$Y_{p,q}$ to be -1 and 1, respectively. Also it is easily seen that
\beq
\G(H)\chi_{p,q}^m=2(m+1)\chi_{p,q}^m,~~~~\G(H')\chi_{p,q}^m=2(q+
\frac{1}{2})\chi_{p,q}^m.
\eeq
It follows that $\chi_{p,q}^m$ has highest weight $(p+1/2,q+1/2)$
for Case 1 (where $m_{\rm max}=p-1/2$) and highest weight
$(p-1/2,q+1/2)$ for Case 2 (where $m_{\rm max}=p-3/2$). This justifies
the use of notation, $|p+\frac{1}{2},m;q+\frac{1}{2}>$ and
$|p-\frac{1}{2},m;q+\frac{1}{2}>$, for these two multiplets, respectively.

Summarizing, we have the following four $sl(2)\oplus u(1)$  multiplets
which span finite-dimensional representations of $osp(2|2)$:
\beqa
|p,m;q>&=&\lt(a^\dg\rt)^{p-m}|0>,~~~m=p,p-1,\cdots, -p,~~~p\geq 0\n
|p-\frac{1}{2},m;q+\frac{1}{2}>&=&\lt(\a_2^\dg+\frac{1}{2}\a_1^\dg
   a^\dg\rt) \lt(a^\dg\rt)^{p-3/2-m}|0>,\n
& &   m=p-\frac{3}{2}, p-\frac{5}{2},\cdots,-(p+\frac{1}{2}),~~~
   p\geq \frac{1}{2},\n
|p+\frac{1}{2},m;q+\frac{1}{2}>&=&\lt(p+m+\frac{3}{2}\rt)\a_1^\dg
   \lt(a^\dg\rt)^{p-1/2-m}|0>\n
& &   -\lt(p-m-\frac{1}{2}\rt)\lt(\a_2^\dg-
   \frac{1}{2}\a_1^\dg a^\dg\rt)\lt(a^\dg\rt)^{p-m-3/2}|0>,\n
& &  m=p-\frac{1}{2}, p-\frac{3}{2},\cdots,-(p+\frac{3}{2}),~~~p\geq 0,\n
|p,m;q+1>&=&\a_1^\dg\a_2^\dg\lt(a^\dg\rt)^{p-2-m}|0>,~~~
   m=p-2, p-3,\cdots,-(p+2),~~~p\geq 0.\n\label{multiplets}
\eeqa
We remark that the trivial 1-dimensional representation (for which
$p=0=q$) is provided by $|0>$ and is not included in the above
expressions.

The actions of the odd generators on these $sl(2)\oplus u(1)$ multiplets
are given by
\beqa
\G(e)|p,m;q>&=&0,\n
\G(f)|p,m;q>&=&\frac{p-m}{2p+1}(q+p+1)
    |p-\frac{1}{2},m-\frac{1}{2};q+\frac{1}{2}>\n
& &    +\frac{1}{2p+1}(q-p)|p+\frac{1}{2},m-\frac{1}{2};q+\frac{1}{2}>,\n
\G(\bar{e})|p,m;q>&=&0,\n
\G(\bar{f})|p,m;q>&=&-\frac{p+m}{2p+1}(q+p+1)
    |p-\frac{1}{2},m-\frac{3}{2};q+\frac{1}{2}>,\n
& & +\frac{1}{2p+1}(q-p)|p+\frac{1}{2},m-\frac{3}{2};q+\frac{1}{2}>,\n
\eeqa
\beqa
\G(e)|p-\frac{1}{2},m;q+\frac{1}{2}>&=&|p,m+\frac{1}{2};q>,\n
\G(f)|p-\frac{1}{2},m;q+\frac{1}{2}>&=&(q-p)|p,m-\frac{1}{2}; q+1>,\n
\G(\bar{e})|p-\frac{1}{2},m;q+\frac{1}{2}>&=&|p,m+\frac{3}{2};q>,\n
\G(\bar{f})|p-\frac{1}{2},m;q+\frac{1}{2}>&=&
  (q-p)|p,m-\frac{3}{2}; q+1>,
\eeqa 
\beqa
\G(e)|p+\frac{1}{2},m;q+\frac{1}{2}>&=&(p+m+\frac{3}{2})|p,m+\frac{1}{2};q>,\n
\G(f)|p+\frac{1}{2},m;q+\frac{1}{2}>&=&-(p-m-\frac{1}{2})(q+p+1)
   |p,m-\frac{1}{2};q+1>,\n
\G(\bar{e})|p+\frac{1}{2},m;q+\frac{1}{2}>&=&-(p-m-\frac{1}{2})
   |p,m+\frac{3}{2};q>,\n
\G(\bar{f})|p+\frac{1}{2},m;q+\frac{1}{2}>&=&(p+m+\frac{3}{2})(q+p+1)
   |p,m-\frac{3}{2};q+1>,
\eeqa 
and
\beqa
\G(e)|p,m;q+1>&=&\frac{p+m+2}{2p+1}
   |p-\frac{1}{2},m+\frac{1}{2};q+\frac{1}{2}>\n
& &-\frac{1}{2p+1}|p+\frac{1}{2},m+\frac{1}{2};q+\frac{1}{2}>,\n
\G(f)|p,m;q+1>&=&0,\n
\G(\bar{e})|p,m;q+1>&=&-\frac{p-m-2}{2p+1}
   |p-\frac{1}{2},m+\frac{3}{2};q+\frac{1}{2}>\n
& &-\frac{1}{2p+1}|p+\frac{1}{2},m+\frac{3}{2};q+\frac{1}{2}>,\n
\G(\bar{f})|p,m;q+1>&=&0,
\eeqa
Note that both $|p,m;q>$ and $|p,m;q+1>$ have dimension $2p+1$, 
$|p-\frac{1}{2}, m;q+\frac{1}{2}>$ has dimension $2p$ and the
dimension of $|p+\frac{1}{2}, m;q+\frac{1}{2}>$ is $2p+2$. So for $q\neq p,
-p-1$, they constitute irreducible
typical representation of dimension $8p+4$ of $osp(2|2)$.

When $q=p,~-p-1$, the representations become atypical. We have different
types of atypical representations. The Casimir for
such representations vanishes, and yet they are not the trivial
one-dimensional representation. As can be seen from the actions of odd
generators to the $sl(2)\oplus u(1)$ multiplets, for $q=p$, if one
starts with $|p,m;q>$ then $|p+\frac{1}{2},
m;q+\frac{1}{2}>$ and $|p,m;q+1>$ disappear and only $|p,m;q>$ and 
$|p-\frac{1}{2}, m;q+\frac{1}{2}>$ survive. They form irreducible atypical
representation of $osp(2|2)$ of dimension $4p+1~ (p\geq 1/2)$.
Similarly, for $q=-p-1$, $|p-\frac{1}{2}, m;q+\frac{1}{2}>$ and $|p,m;q+1>$
do not appear and only $|p,m;q>$ and $|p+\frac{1}{2}, m;q+\frac{1}{2}>$
remain. They constitute irreducible atypical representation of dimension
$4p+3$. Other types of atypical representations  are not irreducible.
One type of such representations are obtained by starting with $|p,m;q+1>$.
As can be seen from the actions of odd generators, these representations
contain all multiplets and have dimension $8p+4$. They are
not fully reducible and are lowest weight indecomposable Kac modules.

\sect{Primary fields of $osp(2|2)^{(1)}_k$ in the standard basis}

Primary fields are fundamental objects in conformal field theories. 
A primary field $\Psi$ has the following OPE with the energy-momentum 
tensor $T(z)$:    
\beq
T(z) \Psi (w)=\frac{\Delta_{\Psi}}{(z-w)^2}\Psi (w)
+\frac{\p _w \Psi (w)}{z-w}+\ldots,
\eeq
\no where the $\Delta_{\Psi}$ is the conformal dimension of $\Psi$. Moreover 
the OPEs of $\Psi$ with the affine currents do not contain poles higher
than first order. A special kind of the primary fields is highest 
weight state. 

The current superalgebra $osp(2|2)^{(1)}_k$ in the standard basis can be
written as
\beq
J_{A}(z)J_{B}(w)=k\frac{{\rm str}(AB)}{(z-w)^2}+f_{AB}^C\frac{J_C(w)}
   {z-w},
\eeq
where $J_A(z)$ stands for the current of $osp(2|2)^{(1)}_k$
corresponding to the $osp(2|2)$ generator $A$ and
$f_{AB}^C$ are structure constants related to $osp(2|2)$
generators $A,B$ and $C$, which can be read off from their
(anti-)commutation relations. In the following, we
shall simply use $A(z)$ to denote $J_A(z)$.

Introduce one bosonic $\bt$-$\gm$ pair,  two fermionic $b$-$c$ type
systems and two free scalar fields. These free fields have the following OPEs: 
\beqa
&&\bt (z)\gm (w)=-\gm(z)\bt(w)=\frac{1}{z-w}, ~~~
\ps (z) \psd (w)=\psd(z)\ps(w)=\frac{1}{z-w},\n
&&\psb (z) \psbd (w)=\psbd(z)\psb(w)=\frac{1}{z-w},~~~~
\ph (z) \ph (w)=-\ln (z-w)= \ph ^{\prime} (z)\ph ^{\prime} (w).\n
\eeqa
Then the free field realization of the currents is given by 
\cite{Bow96,Ras98,Ding03}
\beqa
&&E(z)=\bt (z)-\frac{1}{2}\ps (z) \psbd (z), \n
&&F(z)=-i2\a_{+}\p \ph (z) \gm (z) -\bt (z)\gm ^2 (z) -\psb (z) \psd (z) 
+\frac{1}{2}\gm(z) (\ps (z) \psd (z) - \psb(z) \psbd (z) ) \n
&&~~~~~~~~~~-\frac{1}{4}\gm ^2 (z)\ps (z) \psbd (z) 
+(k-\frac{1}{2}) \p \gm (z),\n
&&H(z)=i2\a_{+}\p \ph (z)-2\bt (z)\gm (z)+ \ps (z) \psd (z) 
-\psb (z) \psbd (z),\n
&&H^{\prime}(z)=2\a _{+} \p \ph ^{\prime} (z)
+ \ps (z) \psd (z) + \psb(z) \psbd (z),\n
&&e(z)=\psd (z)+\frac{1}{2}\gm (z) \psbd (z), \n
&&f(z)=-\a_{+} (i\p \ph (z)-\p \ph ^{\prime} (z))\ps (z)
 +\bt (z) \psb (z) +\frac{1}{2}\bt (z)\gm (z) \ps (z) \n
&&~~~~~~~~~+\frac{1}{2} \psb(z) \psbd (z) \ps (z) +(k+\frac{1}{2})\p \ps (z),\n
&&\bar{e}(z)=\psbd (z), \n
&&\bar{f}(z)=-\a _{+}(i\p \ph (z) +\p \ph ^{\prime} (z) )\psb (z)
-\frac{1}{2}\a _{+}(3i\p \ph (z)-\p \ph ^{\prime} (z) )\gm (z) \ps (z) \n 
&&~~~~~~~~~+\bt (z)\gm (z)\psb (z) - \ps (z) \psd (z) \psb(z)
+\frac{1}{2}\bt (z)\gm ^2 (z) \ps (z) \n
&&~~~~~~~~~-k \p \psb (z) -\frac{1}{2}(k-1)\ps (z) \p \gm (z)
+\frac{1}{2}(k+1)\gm (z) \p \ps (z),\label{free-st}
\eeqa  
where $\a_+ =\sqrt{\frac{k+1}{2}}$, and normal ordering is implied 
in the expressions. In terms of the free fields, the energy-momentum tensor
in the standard basis reads
\beqa
T(z)&=&\bt(z)\gm(z)-\psd (z)\p\ps (z)-\psbd (z)\p\psb(z)\n
& & +\frac{1}{2}\lt(
  [i\p\phi(z)]^2+[\p\phi'(z)]^2\rt)-\frac{1}{\a_+}\lt(i\p^2\phi(z)-
 \p^2\phi(z)\rt).
\eeqa

Now we construct primary fields of the $osp(2|2)$ CFT in the standard
basis.  It is easy to see that the field 
\beq
V_{p,q}(z)={\rm exp} \{\frac{1}{\a _+}(p i \ph (z)- q \ph ^{\prime}(z))\},
\eeq
where $p$ is a positive integer or half-integer and $q$ an arbitrary
complex number which specify the
representation, is a highest weight state of the $osp(2|2)$ current 
superalgebra. The conformal dimension of this field is 
\beq
\Delta_{p,q}=\frac{p(p+1)-q(q+1)}{k+1}.
\eeq
If $q\neq p, -p-1$, then $\Delta_{p,q}\neq 0$ and the corresponding
representations are typical. When $q=p, -p-1$, we have $\Delta_{p,q}=0$
and atypical representations arise. From (\ref{multiplets}), one can
show that the full set of primary fields labelled by $p,q$ are given by
\beqa
S^m _{p,q}(z)&=&\gm (z)^{p-m}V_{p,q}(z), 
~~m=p,~p-1,\cdots, -(p-1),-p, ~~~p\geq 0,\n
s^n _{p,q}(z)&=&\gm (z)^{(p-3/2)-n}
\left(\psb (z)+\frac{1}{2}\gm (z) \ps (z) \right)V_{p,q}(z),\n 
& &n=(p-3/2),\cdots,-(p+1/2),~~~p\geq 1/2, \n
{\tl s}^l _{p,q}(z)&=&\gm (z)^{(p-3/2)-l}
\left((p+l+\frac{3}{2})\gm(z)\psi(z)-(p-l-\frac{1}{2})
  [\psb (z)-\frac{1}{2}\gm (z) \ps (z)] \right)V_{p,q}(z), \n
& &l=(p-1/2),\cdots,-(p+3/2),~~~p\geq 0,\n
{\cal S}^s _{p,q}(z)&=&\gm (z)^{(p-2)-s} \ps (z) \psb (z) V_{p,q}(z),~~
s=(p-2),\cdots,-(p+2),~~~p\geq 0. 
\eeqa 

The dimension of both $S^m _{p,q}(z)$ and 
${\cal S}^s _{p,q}(z)$ is $2p+1$,  ${\tl s}^{l} _{p,q}(z)$ has
dimension $2p$ and ${\tl s}^l _{p,q}(z)$ is $(2p+2)$-dimensional.
So when $q\not= p, -p-1$ the primary fields form an irreducible typical
representation of $osp(2|2)$ of dimension $8p+4$. For irreducible atypical
representation corresponding to $q= p$, 
$S^m _{p,q}(z)$ and $s^n _{p,q}(z)$ are the only non-vanishing fields
and the dimension of the representation is $4p+1~(p\geq
1/2)$. For irreducible atypical representation with $q=-p-1$, 
only $S^m _{p,q}(z)$ and
${\tl s}^l _{p,q}(z)$ survive and the representation is $(4p+3)$-dimensional.

By means of the free field representations, we may
compute the OPEs of $osp(2|2)$ currents with the primary fields.
The results are 
\beqa
&&E (z) S^m _{p,q}(w)= \frac{p-m}{z-w}S^{m+1} _{p,q}(w), \n
&&F (z) S^m _{p,q}(w)= \frac{p+m}{z-w}S^{m-1} _{p,q}(w), \n
&&H (z) S^m _{p,q}(w)= \frac{2m}{z-w}S^{m} _{p,q}(w), \n
&&H^{\prime} (z) S^m _{p,q}(w)= \frac{2q}{z-w}S^{m} _{p,q}(w), \n
&&e (z) S^m _{p,q}(w)=0, \n
&&{\bar e} (z) S^m _{p,q}(w)=0, \n
&&f (z) S^m _{p,q}(w)= \frac{1}{z-w}
\left( \frac{p-m}{2p+1}(q+p+1)s^{m-1/2} _{p,q}(w)
+\frac{1}{2p+1}(q-p){\tl s}^{m-1/2} _{p,q}(w)\right), \n
&&{\bar f }(z) S^m _{p,q}(w)= \frac{1}{z-w}
\left( -\frac{p+m}{2p+1}(q+p+1)s^{m-3/2} _{p,q}(w)
+\frac{1}{2p+1}(q-p){\tl s}^{m-3/2} _{p,q}(w)\right),\n
\eeqa
\beqa
&&E (z) s^n _{p,q}(w)= \frac{(p-3/2)-n}{z-w}s^{n+1} _{p,q}(w), \n
&&F (z) s^n _{p,q}(w)= \frac{(p+1/2)+n}{z-w}s^{n-1} _{p,q}(w), \n
&&H (z) s^n _{p,q}(w)= \frac{2(n+1)}{z-w}s^{n} _{p,q}(w), \n
&&H^{\prime} (z) s^n _{p,q}(w)= \frac{2(q+1/2)}{z-w}s^{n} _{p,q}(w), \n
&&e (z) s^n _{p,q}(w)=\frac{1}{z-w}S^{n+1/2} _{p,q}(w), \n
&&{\bar e} (z) s^n _{p,q}(w)=\frac{1}{z-w}S^{n+3/2} _{p,q}(w), \n
&&f (z) s^n _{p,q}(w)= \frac{q-p}{z-w}{\cal S}^{n-1/2} _{p,q}(w), \n
&&{\bar f }(z)s^n _{p,q}(w)= \frac{q-p}{z-w} {\cal S}^{n-3/2} _{p,q}(w),
\eeqa
\beqa
&&E (z) {\tl s}^l _{p,q}(w)= \frac{p-l-1/2}{z-w}{\tl s}^{l+1} _{p,q}(w),\n
&&F (z) {\tl s}^l _{p,q}(w)= \frac{(p+3/2)+l}{z-w}{\tl s}^{l-1} _{p,q}(w),\n
&&H (z) {\tl s}^l _{p,q}(w)= \frac{2(l+1)}{z-w}{\tl s}^{l} _{p,q}(w), \n
&&{H^{\prime}} (z) {\tl s}^l _{p,q}(w)= \frac{2(q+1/2)}{z-w}
  {\tl s}^{l} _{p,q}(w), \n
&&e (z) {\tl s}^l _{p,q}(w)=\frac{p+l+3/2}{z-w}S^{l+1/2} _{p,q}(w), \n
&&{\bar e} (z) {\tl s}^l _{p,q}(w)=-\frac{p-l-1/2}{z-w}S^{l+3/2} _{p,q}(w),\n
&&f (z) {\tl s}^l _{p,q}(w)= -\frac{p-l-1/2}{z-w}(q+p+1)
  {\cal S}^{l-1/2} _{p,q}(w), \n
&&{\bar f }(z) {\tl s}^l _{p,q}(w)= \frac{p+l+3/2}{z-w}(q+p+1)
  {\cal S}^{l-3/2} _{p,q}(w),
\eeqa
\beqa
&&E (z) {\cal S}^s _{p,q}(w)
= \frac{(p-2)-s}{z-w}{\cal S}^{s+1} _{p,q}(w), \n
&&F (z) {\cal S}^s _{p,q}(w)
= \frac{(p+2)+s}{z-w}{\cal S}^{s-1} _{p,q}(w), \n
&&H (z) {\cal S}^s _{p,q}(w)
= \frac{2(s+2)}{z-w}{\cal S}^{s} _{p,q}(w), \n
&&{H^{\prime}} (z) {\cal S}^s _{p,q}(w)
= \frac{2(q+1)}{z-w}{\cal S}^{s} _{p,q}(w), \n
&&e (z) {\cal S}^s _{p,q}(w)
=\frac{1}{z-w}\lt(\frac{p+s+2}{2p+1}s^{s+1/2} _{p,q}(w)-
  \frac{1}{2p+1}{\tl s}^{s+1/2} _{p,q}(w)\rt), \n
&&{\bar e} (z) {\cal S}^s _{p,q}(w)
=-\frac{1}{z-w}\left(\frac{p-s-2}{2p+1}s^{s+3/2} _{p,q}(w)+
 \frac{1}{2p+1}{\tl s}^{s+3/2} _{p,q}(w)\rt),\n
&&f (z) {\cal S}^s _{p,q}(w)=0, \n 
&&{\bar f }(z) {\cal S}^s _{p,q}(w)= 0.
\eeqa

\sect{Conclusions}

We have studied the representations of $osp(2|2)$ in the standard basis by
means of the coherent state method. All finite-dimensional typical and
atypical representations in this basis have been constructed explicitly. 
In doing so, we have obtained a free boson-fermion 
realization of the $osp(2|2)$ algebra. We have also investigated 
the CFT  associated with the current
superalgebra $osp(2|2)^{(1)}_k$ in the standard basis. We construct 
all primary fields corresponding to finite-dimensional representations
in the standard basis. The CFT is non-unitary as there exists an infinite
family of negative dimensional primary operators in the theory. 
The procedure presented here for the explicit construction of representations
and primary fields may be generalized to other superalgebras. 

The coherent state method is also useful in dealing with
infinite-dimensional representations of current superalgebras. For
example, the boson-fermion realization (\ref{b-f}) could be used to
obtain the free field realization (\ref{free-st}) of the $osp(2|2)$ currents. 
This free field realization gives rise to (reducible)
infinite-dimensional representations of $osp(2|2)^{(1)}_k$ for $k\neq
0$. Irreducible representations may then be obtained by means of the
BRST cohomological analysis, which is out of the scope of the present paper.

\vskip.3in

\no {\bf Acknowledgments:}

This work is financially supported by Australian Research Council.

\bebb{99}

\bbit{Roz92} 
L. Rozanski and H. Saleur, \npb {376} {1992} {461}.

\bbit{Isi94}
J. M. Isidro and A. V. Ramallo, \npb {414}{1994}{715}. 

\bbit{Flohr01}
M. Flohr, preprint hep-th/0111228.

\bbit{Efe83}
K. Efetov, Adv. Phys. {\bf 32}, (1983) 53. 

\bbit{Ber95}
D. Bernard, preprint hep-th/9509137.

\bbit{Mud96}
C. Mudry, C. Chamon and X.-G. Wen, \npb {466} {1996} 383.

\bbit{Maa97}
Z. Maassarani and D. Serban, \npb {489} {1997} {603}.

\bbit{Bas00}
Z.S. Bassi and A. LeClair, \npb {578} {2000} {577}.

\bbit{Gur00}
S. Guruswamy, A. LeClair and A.W.W. Ludwig, \npb {583} {2000} {475}.

\bbit{Lud00}
A.W.W. Ludwig, preprint cond-mat/0012189.

\bbit{Bha01}
M. J. Bhaseen,  J.-S. Caux ,  I. I. Kogan  and  A. M. Tsvelik, 
\npb {618}{2001}{465}.

\bbit{Sch77} 
M. Scheunert, W. Nahm and V. Rittenberg, \jmp {18} {1977} {155};
\jmp {18} {1977} {146}.

\bbit{Mar80} 
M. Marcu, \jmp {21} {1980} {1277}; \jmp {21} {1980} {1284}.

\bbit{Dob86}
V.K. Dobrev, in {\it ``Espoo 1986, Proceedings, Topological and
Geometrical Methods in Field Theory"}, pp.93.

\bbit{Kob94}
K. Kobayashi, Z. Phys. {\bf C61}, (1994) 105.

\bbit{Bow97}
P. Bowcock and A. Taormina, Commun. Math. Phys. {\bf 185}, (1997) 467.

\bbit{Sem97}
A.M. Semikhatov, Theor. Math. Phys. {\bf 112}, (1997) 949.

\bbit{Ding03}
X.M. Ding, M. D. Gould, C. J. Mewton and Y. Z. Zhang, 
 J. Phys. {\bf A36}, (2003) 7649.

\bbit{Bow96}
P. Bowcock, R-L.K. Koktava and A. Taormina, \plb {388} {1996} {303}.

\bbit{Ras98}
J. Rasmussen, \npb {510} {1998} {688}.

\eebb

\end{document}